\documentclass[preprint,superscriptaddress,onecolumn,pre]{revtex4-1}
\usepackage[german,english]{babel} 
\usepackage[export]{adjustbox}
\usepackage{lineno}\usepackage[normalem]{ulem}
\usepackage{booktabs} 
\usepackage{comment} 
\usepackage[utf8]{inputenc} 
\usepackage[T1]{fontenc} 
\usepackage{amsmath,amsfonts,amsthm,amssymb} 
\usepackage{xcolor}
\usepackage{tikz} 
\usepackage{tikz-cd}
\usetikzlibrary{positioning,arrows} 
\usetikzlibrary{decorations.pathreplacing} 
\usepackage{tikzsymbols} 
\usepackage{geometry} 
\usepackage{setspace}
\usepackage[T1]{fontenc} 
\usepackage[utf8]{inputenc} 
\usepackage{har2nat}
\usepackage{multirow}

\newcommand{\bra}[1]{\left(#1\right)}
\newcommand{\Bra}[1]{\left[#1\right]}
\newcommand{\BRA}[1]{\left\{#1\right\}}
\newcommand{\parf}[2]{\frac{\partial #1}{\partial #2}}
\newcommand{\parfD}[2]{\frac{\partial^2 #1}{\partial #2}}
\newcommand{\parfT}[2]{\frac{\partial^3 #1}{\partial #2}}

\newcommand{\dx}{\text{d}\textbf{x}}

\def\ky{k_y^\text{max}}

\begin{document}
	
	\title{Bending and pinching of three-phase stripes: From secondary instabilities to morphological deformations in organic photovoltaics}
	
	\author{Alon Z. Shapira}
	\affiliation{Swiss Institute for Dryland Environmental and Energy Research, Blaustein Institutes for Desert Research, Ben-Gurion University of the Negev, Sede Boqer Campus, Midreshet Ben-Gurion 8499000, Israel}
	
	\author{Nir Gavish}
	\affiliation{Department of Mathematics, Technion - IIT, Haifa, 3200003, Israel}
	
	\author{Hannes Uecker}
	\affiliation{Institute for Mathematics, Carl von Ossietzky University of Oldenburg, P.F 2503, 26111 Oldenburg, Germany}
	
	\author{Arik Yochelis}\email{yochelis@bgu.ac.il}
	\affiliation{Department of Solar Energy and Environmental Physics, Blaustein Institutes for Desert Research, Ben-Gurion University of the Negev, Sede Boqer Campus, Midreshet Ben-Gurion 8499000, Israel}
	\affiliation{Department of Physics, Ben-Gurion University of the Negev, Beer Sheva 8410501, Israel}
	
	\date{\small \today}

	\begin{abstract}
		Optimizing the properties of the mosaic morphology of bulk heterojunction (BHJ) organic photovoltaics (OPV) is not only challenging technologically but also intriguing from the mechanistic point of view. Among the recent breakthroughs is the identification and utilization of a three-phase (donor/mixed/acceptor) BHJ, where the (intermediate) mixed-phase can inhibit morphological changes, such as phase separation. Using a mean-field approach, we reveal and distinguish, between generic mechanisms that alter through transverse instabilities the evolution of stripes: the bending (zigzag mode) and the pinching (cross-roll mode) of the donor/acceptor domains. The results are summarized in a parameter plane spanned by the mixing energy and illumination, and show that donor-acceptor mixtures with higher mixing energy are more likely to develop pinching under charge-flux boundary conditions. The latter is notorious as it leads to the formation of disconnected domains and hence to loss of charge flux. We believe that these results provide a qualitative road-map for BHJ optimization, using mixed-phase composition and therefore, an essential step toward long-lasting OPV. More broadly, the results are also of relevance to study the coexistence of multiple-phase domains in material science, such as in ion-intercalated rechargeable batteries.
	\end{abstract}
	
	\maketitle 
	
	\section{Introduction}
	
	Organic photovoltaics (OPV) are being subjected to intensive research over the past two decades not only due to their potential advantages as portable and/or lightweight technological devices but also for their intriguing physicochemical mechanisms of operation~\cite{kini2020design,vogelbaum2017recently,bonasera2020tackling,zhou2019all}. At the heart of the OPV is the nano-scale mosaic active layer of electron donor (D) and electron acceptor (A) materials, the so-called \textit{bulk heterojunction} (BHJ)~\cite{HQL:2011,C2JM33645F,GKAB:2012,CTA:2011,POLB:POLB23063,KVOWHG:2011,PhysRevLett.108.026601}. This subtle morphology is essential for efficient dissociation of excitons at the D/A interfaces to electrons and holes and for transport of the latter toward the collectors~\cite{schaffer2013direct,schaffer2016morphological,naveed2019interfacial}. The short lifetime of the excitons is translated to a spatial length scale, also known as the \textit{diffusion length}, which respectively sets about tens of nanometer bi-continuous (ideally comb-like) morphology~\cite{mateker2017progress,jorgensen2012stability,CGGHMA:2010,ZSvAMVvM:2009,PhysRevLett.108.026601,AENM:AENM201000023,kouijzer2013predicting,treat2014phase,cardinaletti2014toward,vongsaysy2014formulation,zhou2015phase}.
	
	Recent evidences however, indicate that in some compositions~\cite{ma2014quantification,Reid201227,razzell2013directly,Bartelt2013364,Burke20141923,muller2014active,gasparini2016designing,zhou2019hierarchical,wang2017ternary,zhou2019all} a third phase, which is being referred to as a \textit{mixed-phase} (MP), may additionally become stable along with the pure D/A phases~\cite{liu2014understanding,dkhil2017toward}. This MP has a molecular percolating structure about a 1:1 ratio between the donor and the acceptor molecules~\cite{dkhil2017toward} and thus is distinct from a random distribution although in both cases the averaged quantity is identical. As such, the MP can be thought of as an effective energetic barrier (as being an intermediate metastable state) between the energetically favorable D and A phases~\cite{dkhil2017toward,shapira2019pattern} while keeping the exciton dissociation properties intact. Recent studies indicate that MP plays a role in the evolution of BHJ, ranging from the width and form of the D/A interface~\cite{shapira2019pattern,ma2014quantification,ma2013domain} to an inhibitor of the phase separation process~\cite{dkhil2017toward}. 
	\begin{figure}[bp]
		(a)\includegraphics[height=0.25\textheight,valign=t]{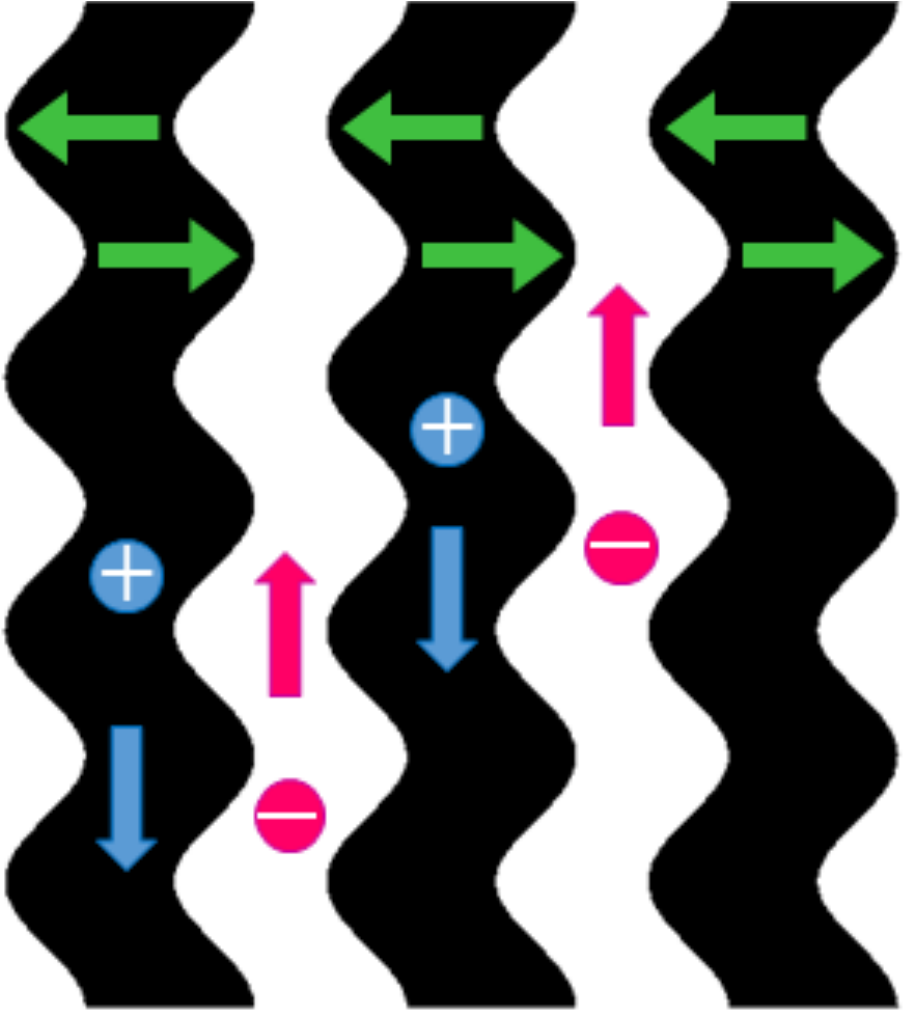} \quad
		(b)\includegraphics[height=0.248\textheight,valign=t]{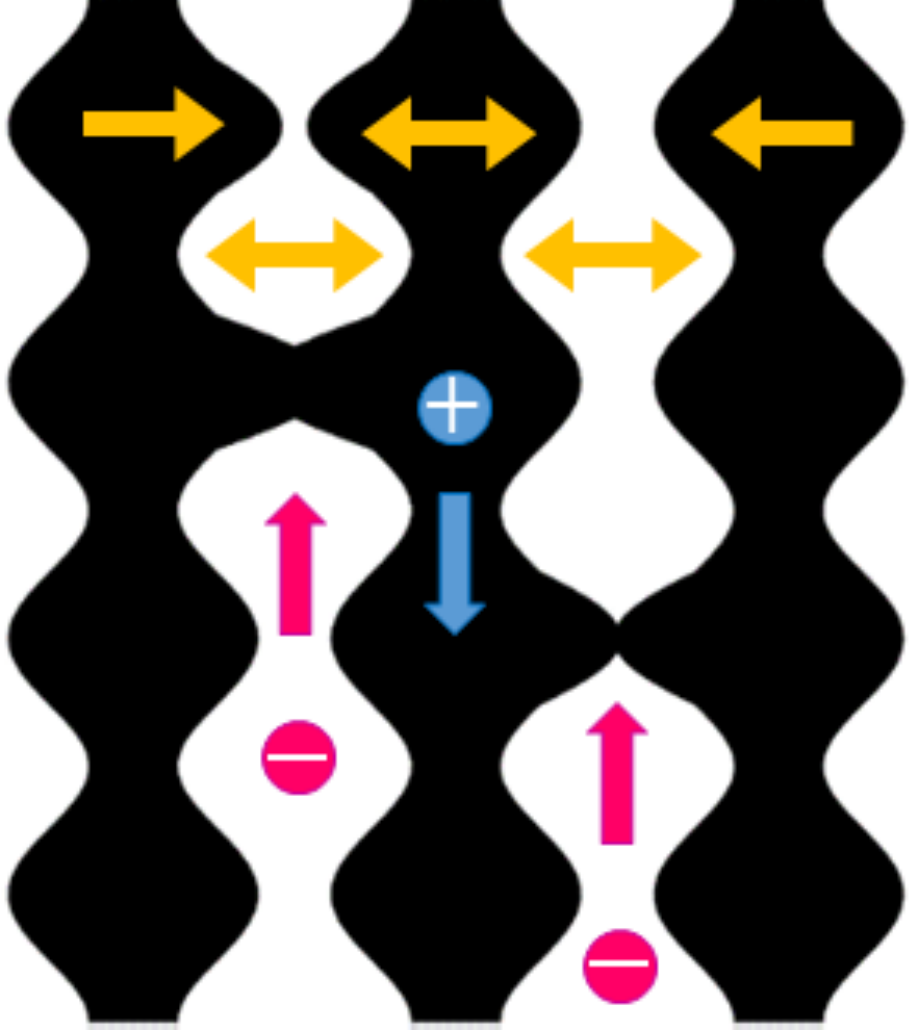}
		\caption[Illustration of bending (ZZ) and pinching (CR) instabilities]{(a) Schematic illustrations of (a) the bending (ZZ) and (b) the cross-roll (CR) instabilities. Horizontal arrows indicate the bending direction while vertical arrows show the directions of the opposite charge fluxes to the electrodes.
		}\label{fig:pre.zzcr}
	\end{figure}
	
	Motivated by three-phase OPV experiments, we study how the intermediate mixed-phase may affect transverse instabilities of striped BHJ by distinguishing between two generic modes and the respective role of boundary conditions (BC): the bending (zigzag) mode and the pinching (cross-roll) mode that is critical for operation since it destroys flux of charges to collectors, as schematically demonstrated in Fig.~\ref{fig:pre.zzcr}. We use a recently proposed Shapira-Gavish-Yochelis mean-field model~\cite{shapira2019pattern} that incorporates the morphological evolution of a three-phase BHJ under illumination and for analysis we employ the generalized eigenvalue methodology~\cite{gavish2017spatially,shapira2020stripes} to identify the instability onsets. Specifically, we elaborate on {how the} stability of the BHJ to pinching depends on the increase of energetic barrier of the mixing energy, i.e., the depth of the intermediate well in the free energy, and exemplify the results in the parameter plane spanned by well depth 
	{and illumination strength}. The generic nature of the results paves a plausible strategy to control the morphological stability of the BHJ under illumination.
	
	\section{Determining the donor-acceptor ratio}
\begin{figure}[t!]
		\centering 
		(a)\includegraphics[height=0.25\textheight,valign=t]{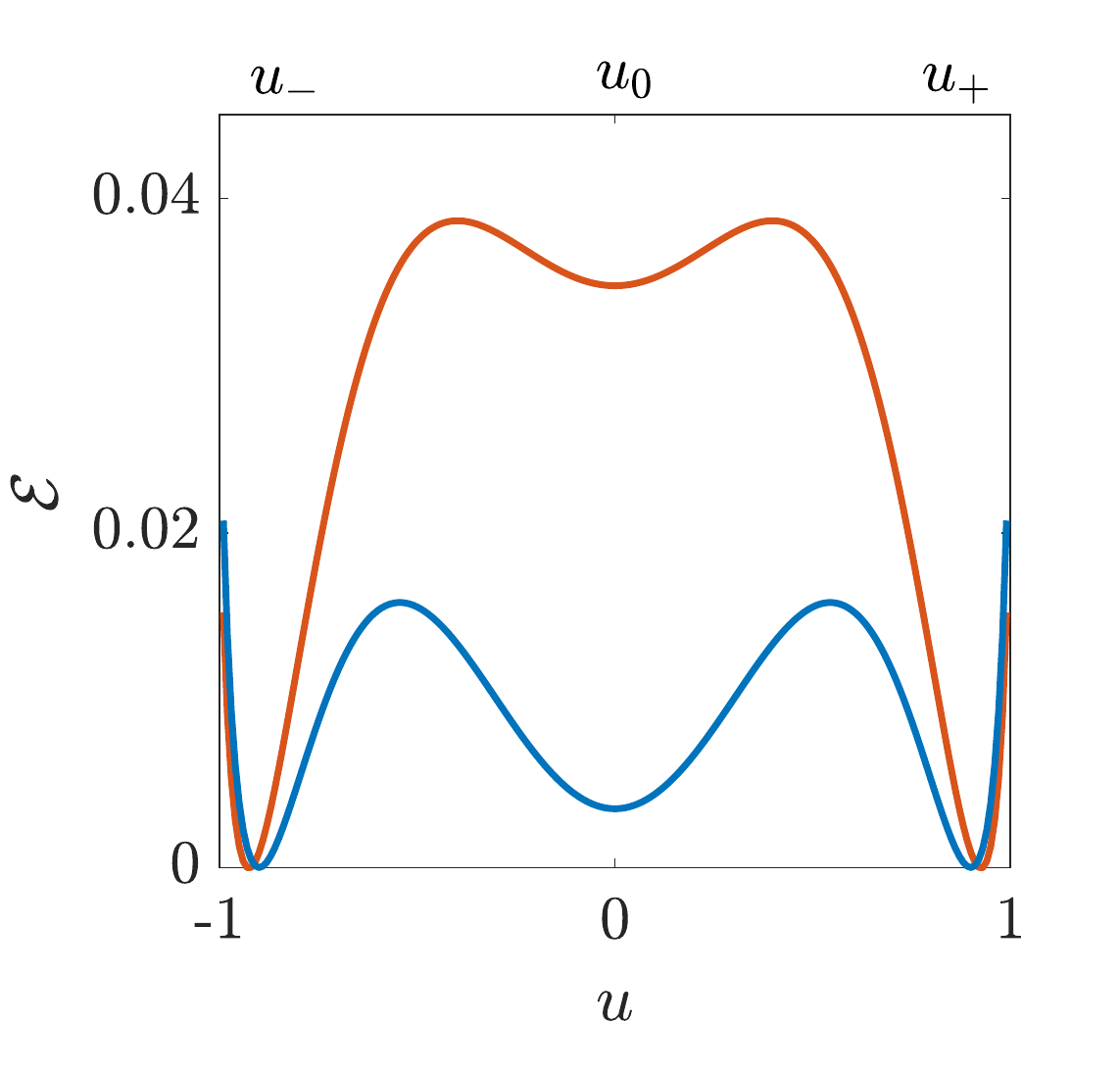}
		(b)\includegraphics[height=0.25\textheight,valign=t]{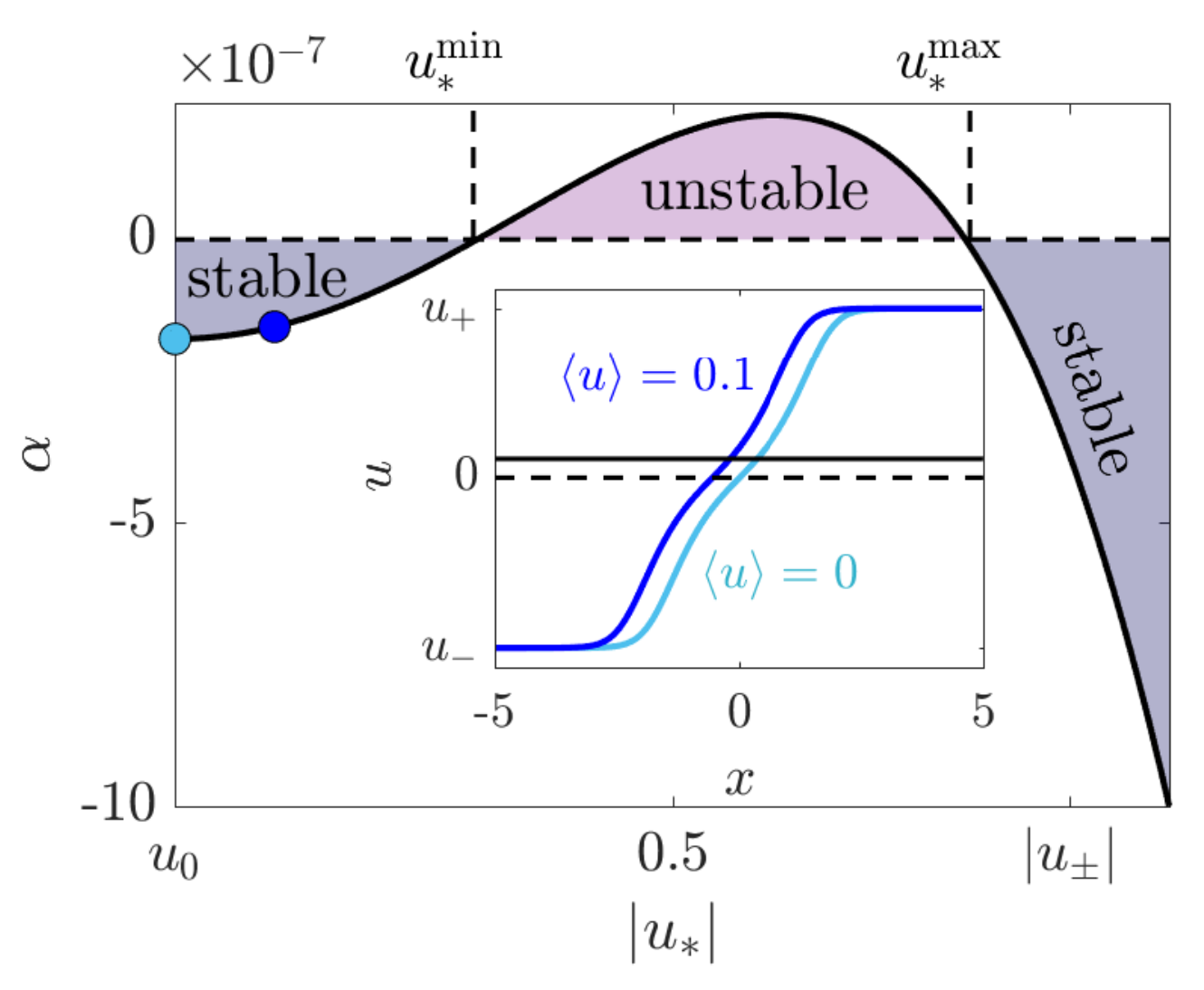}
		\caption[Stable and unstable mass]{(a) Free energy functional 
			\eqref{eq:pre.energy} for uniform $u$, with $\xi=2.65$, $e_0=-0.034$ (blue, {genuine triple well}) and $\xi=2.8$, $e_0=0.028$ (red, {closer to double well}), where~$\xi$ corresponds to the mixing energy. (b) Stability and instability intervals for uniform solutions determined by the critical values $u_*^\text{min}\approx0.30$ and $u_*^\text{max}\approx0.80$. The inset demonstrates three-phase interface solutions, as computed from \eqref{eq:pre.cahnHilliard}, for $\langle u \rangle=0$ and $\langle u \rangle=0.1$ (as marked by `$\bullet$' in the main figure, respectively); the location of the interface with $\langle u \rangle=0$ is at $x=0$. The horizontal solid and dashed lines indicate the value $u_*=\langle u \rangle=0.1$ and $u_*=\langle u \rangle=0$, respectively. Parameters: $D_u=1$, $\beta=0.5$, $\xi=2.65$, $\lambda=0.1$,  $k=10^{-3}$.}\label{fig:pre.ustab}
	\end{figure}

	In the dark, the free energy comprises the entropy and the mixing energy for the material order parameter~\cite{shapira2019pattern}, $u:=\varphi_{\text A}-\varphi_{\text D}\in [-1,1]$, where $\varphi_{\text A},\varphi_{\text D}$ are the respective fractions of the A/D phases. Its dimensionless form reads 
	\begin{equation}\label{eq:pre.energy}
		\mathcal{E}_{\text M}(u)=\int_\Omega \underbrace{\frac{1-u}{2}\ln\frac{1-u}{2}}_\text{donor's entropy}+ \underbrace{\frac{1+u}{2}\ln\frac{1+u}{2}}_\text{acceptor's entropy}+\underbrace{\frac{\beta}{2}(1-u^2)(u^2+\xi)+\frac{\lambda}{2}|\nabla u|^2-e_0}_\text{mixing energy}~\dx,    
	\end{equation}
	where $\Omega$ is the domain, {which we take to be a rectangle 
		$\Omega=(0,l_x)\times (0,l_y)$.} Further, 
	$e_0$ is a reference energy density for which the minimum of $\mathcal{E}_{\text M}(u)$ is zero, $\beta$ determines the ratio between mixing energy and entropy, and $\xi$ determines the depth of the intermediate well such that small $\xi$ corresponds to a lower mixing energy (see Fig.~\ref{fig:pre.ustab}(a)), {and $\lambda$ is the penalty for creation of multiple interfaces and associated with the width of the interface}. Due to entropy, the minimum energy of donor-rich ($u:=u_-$) and acceptor-rich ($u:=u_+$) phases are shifted from $u=\pm1$ to slightly lower values in $|u|$, while the mixed-phase always sits at $u:=u_0=0$, as shown in Fig.~\ref{fig:pre.ustab}(a). Due to conservation of the order parameter, however, there are many other uniform solutions $u=u_*$ and these solutions are related to the D:A ratio of non-uniform solutions, e.g., D-A interfaces. The connection between the $u_*$ and the D:A ratio is made through averaging of $u$ in one space dimension (1D), 
	\begin{equation}\label{eq:avrg}
		\langle u \rangle:=l_x^{-1}\int_0^{l_x} u\,{\text d}x.
	\end{equation}
	For the \textit{symmetric} case $\langle u \rangle=0$, the amount of donor and acceptor is identical so that the interface is located at {$x=l_x/2$,}
	and for the \textit{asymmetric} case, where $|\langle u \rangle|>0$, this location is shifted; note that for the uniform states $\langle u \rangle=u_*$. Thus, for non-uniform solutions that are of interest here, it is required to identify the allowed range of $\langle u \rangle$ and we do it by looking at the stability of $u_*$.
	
	The evolution equation~\cite{shapira2019pattern} (in the dark and with mobility $D_u(1-u^2)$) reads as
	\begin{equation}\label{eq:pre.cahnHilliard}
		\parf{u}{t}=D_u\parfD{u}{x^2}+D_u\frac\partial{\partial x}\BRA{\bra{1-u^2} \Bra{\beta (1-6u^2-\xi)\parf{u}{x}-\lambda\parfT{u}{x^3}}},
	\end{equation}
	where $D_u$ is the diffusion coefficient. Linear stability analysis (performed on an infinite domain) about uniform states $u=u_*$ corresponds to 
	\begin{equation}\label{eq:pre.pertu}
		u-u_*\propto e^{\alpha t +ikx} +\text{c.c.},
	\end{equation}
	where $\text{c.c.}$ is the complex conjugate and $\alpha$ is the perturbation growth rate of wavenumber $k$ and given by
	\begin{equation}\label{eq:pre.alpha}
		\alpha(k)=-D_u k^2\left\{1+(1-u_*^2)
		\left[\beta(1-6u_*^2-\xi)+\lambda k^2\right]\right\}.
	\end{equation}
	The instability of $u=u_*$ is of a typical long-wavenumber type~\cite{CrossHohenberg1993} and the regime of unstable steady state solutions $u^{\min}_*<|u_*|<u^{\max}_*$ is obtained by taking the limit $\alpha(k)\to 0$ as $k\to 0$ (see Fig.~\ref{fig:pre.ustab}(b)), where
	\[    
	u_*^\text{min}=\frac{1}{2\sqrt{3}}\sqrt{7-\xi-\sqrt{(5+\xi)^2-24/\beta}},\quad
	u_*^\text{max}=\frac{1}{2\sqrt{3}}\sqrt{7-\xi+\sqrt{(5+\xi)^2-24/\beta}}.
	\]
	These results imply that any choice within the $u^{\min}_*<|u_*|<u^{\max}_*$ range will result in a coarsening dynamics that minimizes the energetic penalty of interfaces, following which phase separation is achieved (not shown here). In the inset of Fig.~\ref{fig:pre.ustab}(b), we show that indeed the interface solutions are bi-asymptotic to $u_\pm$ and that the location of the interface shifts according to $\langle u \rangle$ that is set by $u_*$. Thus, in the analysis that follows, we will focus on the regime $0<|u_*|<u^{\min}_*$, 
	{under illumination.} 
	
	\section{Existence of stripes under illumination and the effect of D-A asymmetry}
	
	Illumination {leads to the creation of} excitons which dissociate at the D/A interface and thus, drive the BHJ out of equilibrium. The (dimensionless) total free energy under illumination takes the form~\cite{shapira2019pattern}
	\begin{equation}\label{eq:pre.freeEnergy}
	\nonumber \mathcal{E}=\mathcal{E}_{\text M}+\int_\Omega \underbrace{\chi\ln \chi +p\ln p+n\ln n}_\text{charges entropy} + \underbrace{\phi(p-n)-\frac{\epsilon}{2}|\nabla\phi|^2}_\text{electrostatic energy} +\underbrace{\frac{\zeta}{2}\big{[} 	p\left(1+u\right)^2+n\left(1-u\right)^2\big{]}}_\text{charge affinity} \dx,
	\end{equation}
	{where the corresponding equations of motion also} incorporate the generation/recombination following~\citet{Buxton2006}, and fluxes of electrical charges coupled to morphological evolution of the BHJ order parameter~\cite{shapira2019pattern}:
	\begin{subequations}  \label{eq:pre.model}
		\begin{align}
			\parf{u}{t}=&\underbrace{D_u{\nabla^2 u}+D_u\nabla\cdot\BRA{\bra{1-u^2} \Bra{\beta (1-6u^2-\xi)\nabla u-\lambda\nabla^3u}}}_{\rm phase~separation}\nonumber\\
			&+\underbrace{D_u \zeta\nabla\cdot \BRA{ \bra{1-u^2} \Bra{\bra{p+n}\nabla u+\bra{1+u}\nabla p-\bra{1-u}\nabla n}}}_{\rm donor/acceptor~affinity~to~charges}
			\label{eq:pre.u}, \\
			\parf{\chi}{t}=&\underbrace{\nabla^2 \chi}_{\rm diffusion}-\underbrace{\tau^{-1}\bra{1-u^2}\chi}_{\rm dissociation}-\underbrace{\chi}_{\rm decay}\label{eq:pre.chi}+\underbrace{G}_{\rm generation}, \\
			\parf{p}{t}=&D_p\nabla\cdot[\underbrace{p\nabla\phi+\nabla p}_{\rm drift-diffusion}+\underbrace{\zeta p(1+u)\nabla u}_{\rm charge~ affinity}]+\tau^{-1}\bra{1-u^2}\chi-\underbrace{\gamma\,np}_{\rm recombination}, \label{eq:pre.p} \\
			\parf{n}{t}=&D_n\nabla\cdot\Bra{-n\nabla\phi+\nabla n-\zeta n(1-u)\nabla u}+\tau^{-1}\bra{1-u^2}\chi-\gamma\,np, \label{eq:pre.n}\\
			0=&\nabla \cdot \Bra{\epsilon \nabla \phi}+p-n\label{eq:pre.phi}.
		\end{align}
	\end{subequations}
	Here the fields $\chi,p,n$ stand for excitons, holes and electrons, respectively, $\phi$ is the electric potential, $D_p,D_n$ are the respective diffusion constants, $\zeta$ is the interaction energy between electron/holes and donor/acceptor, $\tau$ is the excitons dissociation time, $G$ is the 
	excitons generation rate, $\gamma$ is the 
	electron-hole recombination rate, and $\epsilon$ is the permittivity. For details we refer the reader to~\citet{shapira2019pattern}.
	
\begin{figure}[tp]
		(a)\includegraphics[height=0.245\textheight,valign=t]{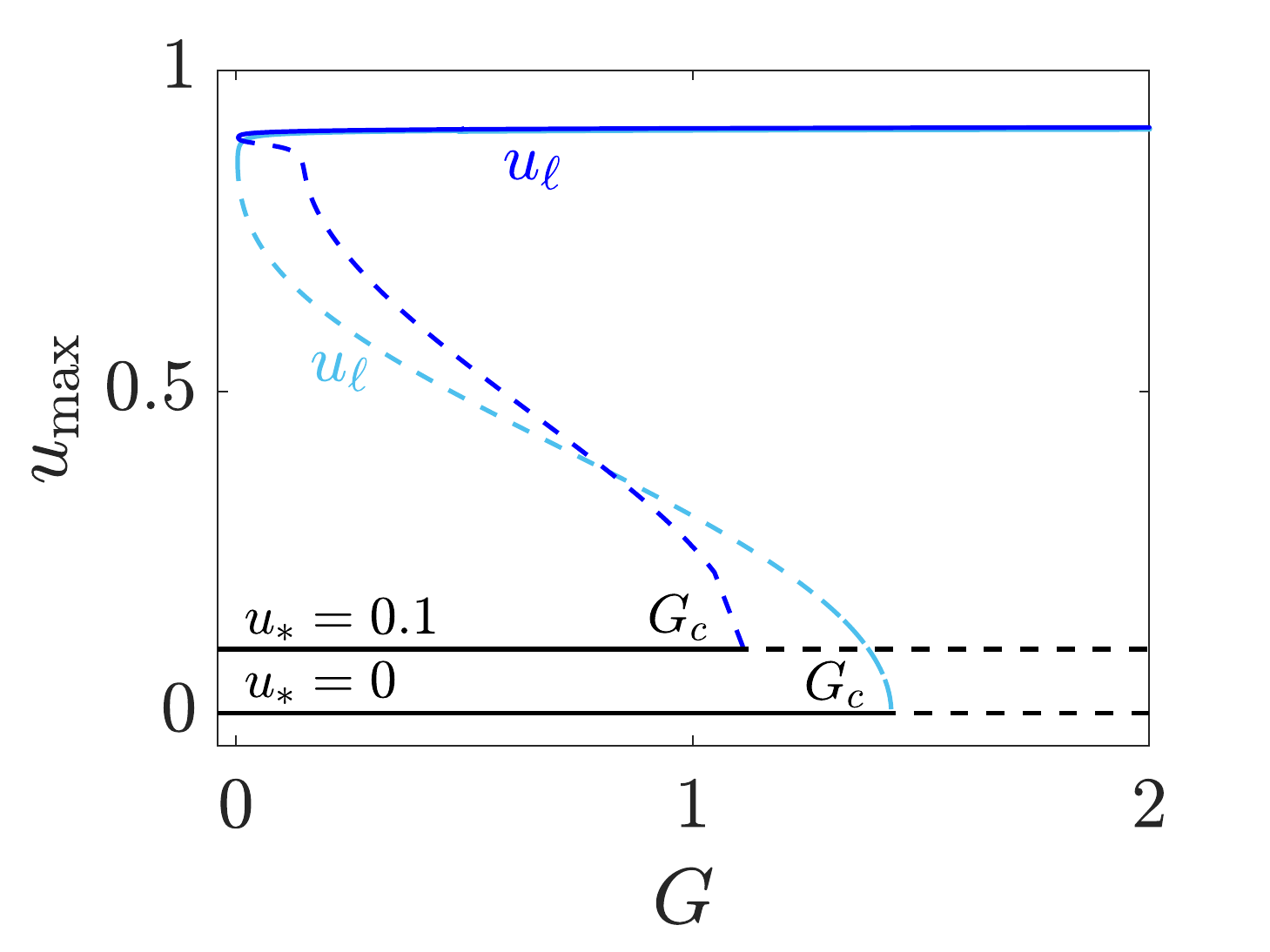}
		(b)\includegraphics[height=0.245\textheight,valign=t]{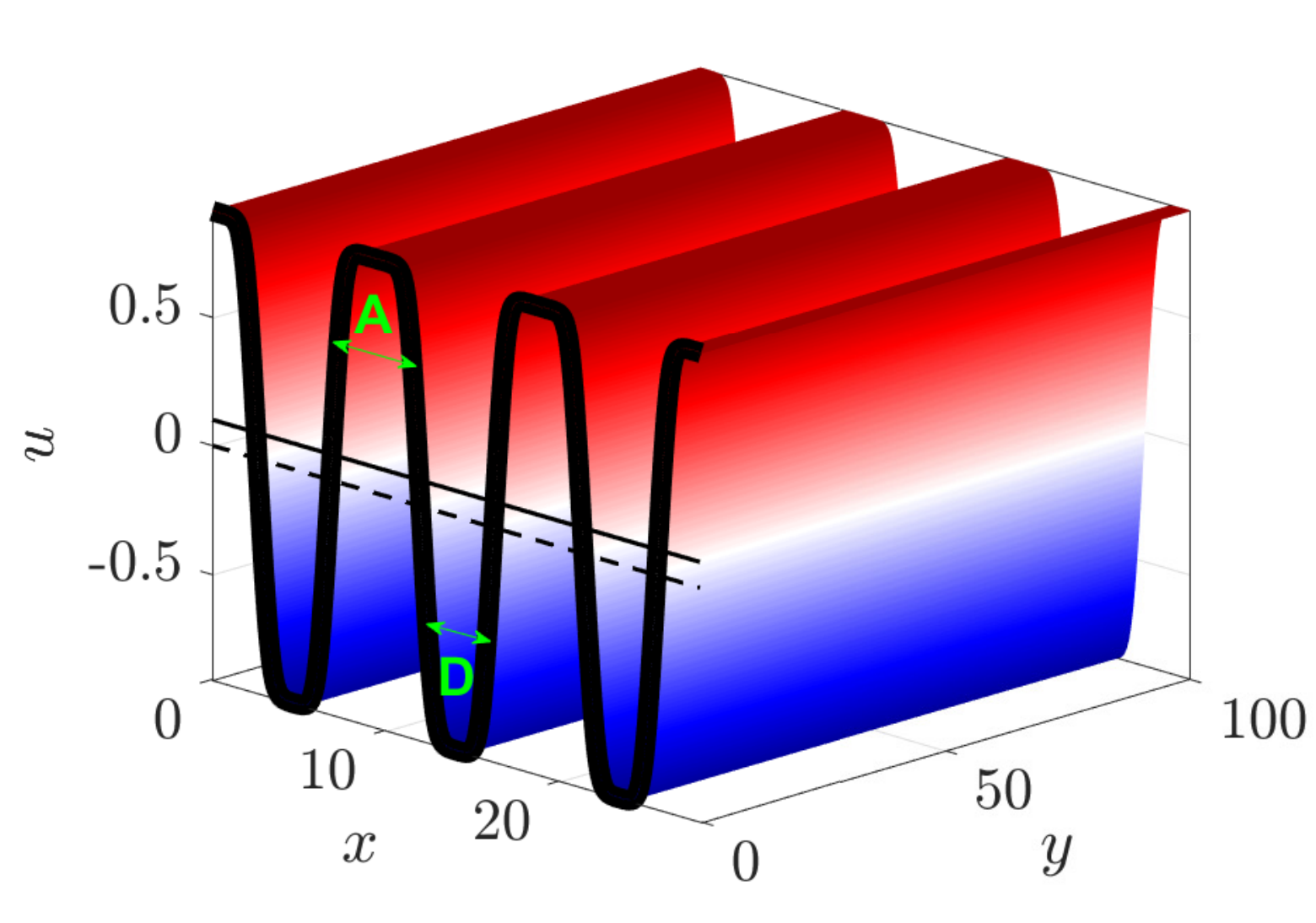}
		\caption[Bifurcation diagram of symmetric and asymmetric solutions]{(a) Bifurcation diagram showing the branches of symmetric (cyan) and asymmetric (blue) periodic solutions in 1D, $u_\ell$, computed for system~\eqref{eq:pre.model}; solid/dashed lines indicate stable/unstable solutions. For these computations we employed the numerical continuation package pde2path~\cite{uecker2014numerical,dohnalpde2path}, {with periodic BC}. 
			The periodic solutions bifurcate from the uniform solutions $u_*=0$ and $u_*=0.1$ at the respective values of $G=G_c$. (b) Asymmetric stripe pattern plotted at $G=2$ after extending $u_\ell$ in the $y$ direction. The solid horizontal line indicates $\langle u \rangle=0.1$. The arrows at $u=\pm0.5$ indicate the width of the acceptor (A) and donor (D) phases, with A being wider. The domain size is $\Omega=[0,3\ell_c]\times[0,100]$ and the critical wavelength of $u_\ell$ is $\ell_c \approx9.38$. Parameters: $\beta=0.5$, $\xi=2.65$, $\lambda=0.1$, $\zeta=4$, $\epsilon=0.25$, $\tau=\gamma=100$, $D_p=D_n=3$, $D_u=1$.}\label{fig:pre.bif}
	\end{figure}

	Uniform solutions of system~\eqref{eq:pre.model} are given by $\textbf{U}_*=(u_*,\chi_*,p_*,n_*,0)$, where
	$\chi_*=\tau G/(\tau+1-u_*^2)$ and $p_*=n_*=\sqrt{G(1-u_*^2)/(\gamma(\tau+1-u_*^2))}$. Linear analysis in 1D, by replacing $u(x)$ with $\textbf{U}(x)$, shows that in range $0\leq |u_*|<u_*^{\min}$, the uniform solution $\textbf{U}_*$ goes through a subcritical finite wavenumber instability at $G=G_c$, giving rise to periodic solutions $\textbf{U}_\ell(x)$ with wavenumber $k_c$, that corresponds to the spatial wavelength $\ell_c=2\pi/k_c$, where for $u_*=0$ we get \[G_c=\frac{\epsilon^2\gamma(\tau+1)(\beta\xi-\beta-1)^2}{4(\lambda-\epsilon\zeta+\epsilon\zeta^2-2\zeta\sqrt{\epsilon\lambda})^2},
	\] and 
	\[k^2_c=\frac{(1-\zeta\sqrt{\epsilon/\lambda})(\beta\xi-\beta-1)}{\lambda-\epsilon\zeta+\epsilon\zeta^2-2\zeta\sqrt{\epsilon\lambda}},
	\]
	while for $|u_*|>0$ the critical values are computed numerically. The periodic solutions $\textbf{U}_\ell(x)=(u_\ell(x),\chi_\ell(x),p_\ell(x),n_\ell(x),\phi_\ell(x))$ bifurcate toward the stable portion of $\textbf{U}_*$, that is in direction $G<G_c$, and thus, are initially unstable. Then they grow in amplitude and stabilize after the saddle node bifurcation that is located close to $G=0$, and continue to be stable as $G$ increases, as shown in Fig.~\ref{fig:pre.bif}(a)]. 
	
	Notably, conservation of the order parameter $u$ forces also the periodic solutions $u_\ell$ to keep the average value that is initially set by $u_*$. Namely, periodic solutions (which can be extended in $y$ direction to form stripes) that bifurcate from $u_*=0$ correspond to \textit{symmetric} stripes (i.e., identical width of the donor and the acceptor domains) while periodic solutions that bifurcate from $u_*=0.1$, for example, are \textit{asymmetric}, in which acceptor domains are wider; the latter is demonstrated in \ref{fig:pre.bif}(b). Next, we calculate the stability properties of stripes in the transverse direction.
	
	\section{Transverse instability of three-phase donor/mixed/acceptor stripes}
	
	{For the linear transverse instability of stripes to zigzag (ZZ) that corresponds to bending and to cross-roll (CR) that causes pinching (see Fig.~\ref{fig:pre.zzcr}), we employ a general space dependent eigenvalue method~\cite{greenside1984nonlinear,greenside1985stability,thiele2003front,kolokolnikov2006zigzag,kolokolnikov2006stability,burke2007homoclinic,diez2012instability} that has been used for example in the context of convection rolls, stripes in reaction-diffusion media, and thin fluid films. However, due to application to OPV, our interest here is to reveal the impact of physical boundary conditions on the instability of stripes, i.e., on non-periodic domains in $y$ directions.}
	
	\subsection{Linear analysis on unbounded domains}
	
	{We start however, by performing a general analysis of stripes on non-physical infinite (periodic in $y$ direction) domains~\cite{shapira2020stripes}
		\begin{equation}\label{eq:pre.pert}
			\textbf{U}(t,x,y)-\textbf{U}_\ell(x)\propto \tilde{\textbf{U}}(x)e^{\eta t+ik_yy}+\text{c.c.},
		\end{equation}
		where $\eta$ is the growth rate of the wavenumber, $k_y$, in the transverse direction to $\textbf{U}_\ell(x)$, and $\tilde{\textbf{U}}(x)=(\tilde{u}(x),\tilde{\chi}(x),\tilde{p}(x),\tilde{n}(x),\tilde{\phi}(x))$ is always the periodic eigenfunction. This formulation introduces a generalized eigenvalue system
		\begin{equation}\label{eq:pre.evp}
			\eta\mathcal{M}\tilde{\textbf{U}}=\mathcal{L}\tilde{\textbf{U}}.
		\end{equation}
		In~\eqref{eq:pre.evp} $\mathcal{M}$ is a singular projection matrix~\cite{gavish2017spatially} 
		\begin{equation}
			\mathcal{M}=\left(\begin{array}{ccccc}
				1 & 0 & 0 & 0 & 0\\[-2mm]
				0 & 1 & 0 & 0 & 0\\[-2mm]
				0 & 0 & 1 & 0 & 0\\[-2mm]
				0 & 0 & 0 & 1 & 0\\[-2mm]
				0 & 0 & 0 & 0 & 0
			\end{array}\right),
		\end{equation}
		$\mathcal{L}$ is a linear operator  
		\begin{equation}\label{eq:L}
			\mathcal{L}[\textbf{U}_\ell;k_y]=\left(\begin{array}{ccccc}
				D_u\mathcal{L}_{1,1} & 0 & D_u\mathcal{L}_{1,3} & D_u\mathcal{L}_{1,4} & 0\\
				2\chi u/\tau & \mathcal{L}_{2,2} & 0 & 0 & 0\\
				\mathcal{L}_{3,1} & (1-u^2)/\tau & \mathcal{L}_{3,3} & -\gamma p & \mathcal{L}_{3,5}\\
				\mathcal{L}_{4,1} & (1-u^2)/\tau & -\gamma n & \mathcal{L}_{4,4} & \mathcal{L}_{4,5}\\
				0 & 0 & 1 & -1 & \epsilon(\hat{\partial}^2_x-k_y^2)
			\end{array}\right),
		\end{equation}
		where
		{\small
			$\mathcal{L}_{1,1}=\lambda\big{[}1{-}u^2\big{]}{\partial}^4_x {+} 2\lambda uu_x{\partial}^3_x {+} \mathcal{L}_{1,1}^{(0)} {+} \mathcal{L}_{1,1}^{(1)}\partial_x {+} \mathcal{L}_{1,1}^{(2)}\partial_x^2$, \\ 
			$\displaystyle \mathcal{L}_{1,1}^{(0)}=\big{\{}\begin{array}[t]{l}
				\lambda\big{[}2uu_{xxxx}{+}2u_xu_{xxx}{-}k_y^4(1{-}u^2)\big{]}{+}\beta\big{[}u_{xx}(24u^3{-}14u){+}u_x^2(72u^2{-}14) {-}k_y^2(6u^4{-}7u^2{+}1)\\
				{+}\xi(2u_x^2{+}2uu_{xx}{+}k_y^2(1{-}u^2))\big{]} {+}\zeta\big{[}p_{xx}(1{-}2u{-}3u^2){+}n_{xx}(1{+}2u{-}3u^2)\\
				{-}2(uu_{xx}{+}u_x^2)(p{+}n){+}u_x((1{-}4u)n_x{-}(1{+}4u)p_x){-}k^2(p{+}n)(1{-}u^2)\big{]} {-}k_y^2\big{\}},\end{array}$\\
			$\mathcal{L}_{1,1}^{(1)}=
			\big{\{}2\lambda u\big{[}u_{xxx}{-}k_y^2u_x\big{]}{+}4\beta uu_x\big{[}12u^2{-}7{+}\xi\big{]}{-}\zeta\big{[}4uu_x(p{+}n){+}2p_x(2u^2{-}1{+}u){+}2n_x(2u^2{-}1{-}u)\big{]}\big{\}}$, \\
			$\mathcal{L}_{1,1}^{(2)}=
			\big{\{}1{+}\beta\big{[}6u^4{-}7u^2{+}1{-}\xi(1{-}u^2)\big{]} {+}\zeta(p{+}n)(1{-}u^2) {+}2\lambda k_y^2(1{-}u^2)\big{\}}$,\\
			$\mathcal{L}_{1,3}=\zeta\big{[}(1{+}u)(1{-}u^2)\big{]}{\partial}^2_x {-}2\zeta u_x\big{[}2u^2{+}u{-}1\big{]}{\partial}_x {-}\zeta\big{\{}2uu_x^2{-}u_{xx}(1{-}u^2){+}k_y^2(1{+}u)(1{-}u^2)\big{\}}$,\\
			$\mathcal{L}_{1,4}=\zeta\big{[}(u{-}1)(1{-}u^2)\big{]}{\partial}^2_x {-}2\zeta u_x\big{[}2u^2{-}u{-}1\big{]}{\partial}_x {-}\zeta\big{\{}2uu_x^2{-}u_{xx}(1{-}u^2){+}k_y^2(u{-}1)(1{-}u^2)\big{\}}$,\\ 
			$\mathcal{L}_{2,2}={\partial}^2_x {+} 1{-}k_y^2{+}(1{-}u^2)/\tau$,\\ 
			$\mathcal{L}_{3,1}=D_p\zeta p(1{+}u){\partial}^2_x {+} D_p\zeta\big{[}p_x(1{+}u){+}2pu_x\big{]}{\partial}_x {+} \big{\{}D_p\zeta\big{[}pu_{xx}{+}p_xu_x{-}k_y^2p(1{+}u)\big{]}{-}2\chi u/\tau\big{\}}$,\\ 
			$\mathcal{L}_{3,3}=D_p{\partial}^2_x {+} D_p\big{[}\phi_x{+}\zeta u_x(1{+}u)\big{]}{\partial}_x {+} \big{\{}D_p\big{[}\phi_{xx}{-}k_y^2{+}\zeta[u_{xx}(1{+}u){+}u_x^2]\big{]}{-}\gamma n \big{\}}$,\\ 
			$\mathcal{L}_{3,5}=D_p\big{\{} p\hat{\partial}^2_x {+} p_x\hat{\partial}_x {-}k_y^2p\big{\}}$,\\ 
			$\mathcal{L}_{4,1}=D_n\zeta n(u{-}1){\partial}^2_x {+} D_n\zeta\big{[}n_x(u{-}1){+}2nu_x\big{]}{\partial}_x {+} \big{\{}D_n\zeta\big{[}nu_{xx}{+}n_xu_x{-}k_y^2n(u{-}1)\big{]}{-}2\chi u/\tau\big{\}}$,\\ 
			$\mathcal{L}_{4,4}=D_n{\partial}^2_x {+} D_n\big{[}{-}\phi_x{+}\zeta u_x(u{-}1)\big{]}{\partial}_x {+} \big{\{}D_n\big{[}{-}\phi_{xx}{-}k_y^2{+}\zeta[u_{xx}(u{-}1){+}u_x^2]\big{]}{-}\gamma p \big{\}}$,\\ 
			$\mathcal{L}_{4,5}={-}D_n\big{\{} n\hat{\partial}^2_x {+} n_x\hat{\partial}_x {-}k_y^2n\big{\}}$.}
		\\
		In addition, we employ in~\eqref{eq:L} spatial operators $\partial_x:=\mathcal{G}$, $\partial_x^2:=\mathcal{D}$ with periodic boundary conditions~\cite{numericalRecipes}
		\begin{equation*}
			\hspace{-5mm}\mathcal{G}\approx\frac{1}{2\Delta x}\left(\begin{array}{rrrrrr}
				0 & 1 &  & & & -1\\
				-1 & 0 & 1 &  & &\\
				& -1 & 0 & 1 &  &\\
				&  &  & \ddots &  & \\
				& &  & -1 & 0 & 1\\
				1 & & &  & -1 & 0
			\end{array}\right),\quad
			\mathcal{D}\approx\frac{1}{\Delta x^2}\left(\begin{array}{rrrrrr}
				-2 & 1 &  & & & 1\\
				1 & -2 & 1 &  & &\\
				& 1 & -2 & 1 &  &\\
				&  &  & \ddots &  & \\
				& &  & 1 & -2 & 1\\
				1 & & &  & 1 & -2
			\end{array}\right),
		\end{equation*}
		where empty entries are zeros and $\Delta x$ is the spatial distance between two points on the uniform grid and respectively, the operators $\hat{\partial}_x:=\hat{\mathcal{G}}$, $\hat{\partial}_x^2:=\hat{\mathcal{D}}$ with two-sided homogeneous Dirichlet boundary conditions to eliminate potential jumps
		\begin{equation*}
			\hspace{-5mm}\hat{\mathcal{G}}\approx\frac{1}{2\Delta x}\left(\begin{array}{rrrrrr}
				0 & 1 &  & & & \\
				-1 & 0 & 1 &  & &\\
				& -1 & 0 & 1 &  &\\
				&  &  & \ddots &  & \\
				& &  & -1 & 0 & 1\\
				& & &  & -1 & 0
			\end{array}\right),\quad
			\hat{\mathcal{D}}\approx\frac{1}{\Delta x^2}\left(\begin{array}{rrrrrr}
				-2 & 1 &  & & & \\
				1 & -2 & 1 &  & &\\
				& 1 & -2 & 1 &  &\\
				&  &  & \ddots &  & \\
				& &  & 1 & -2 & 1\\
				& & &  & 1 & -2
			\end{array}\right),
		\end{equation*}
		where, for higher-order derivatives we used the identities $\partial_x^3=\mathcal{G}\mathcal{D}$ and $\partial_x^4=\mathcal{D}^2$.}
	\begin{figure}[t!]
		\begin{tabular}{ll}
			(a)\hspace{3mm}\includegraphics[height=0.18\textheight,valign=t]{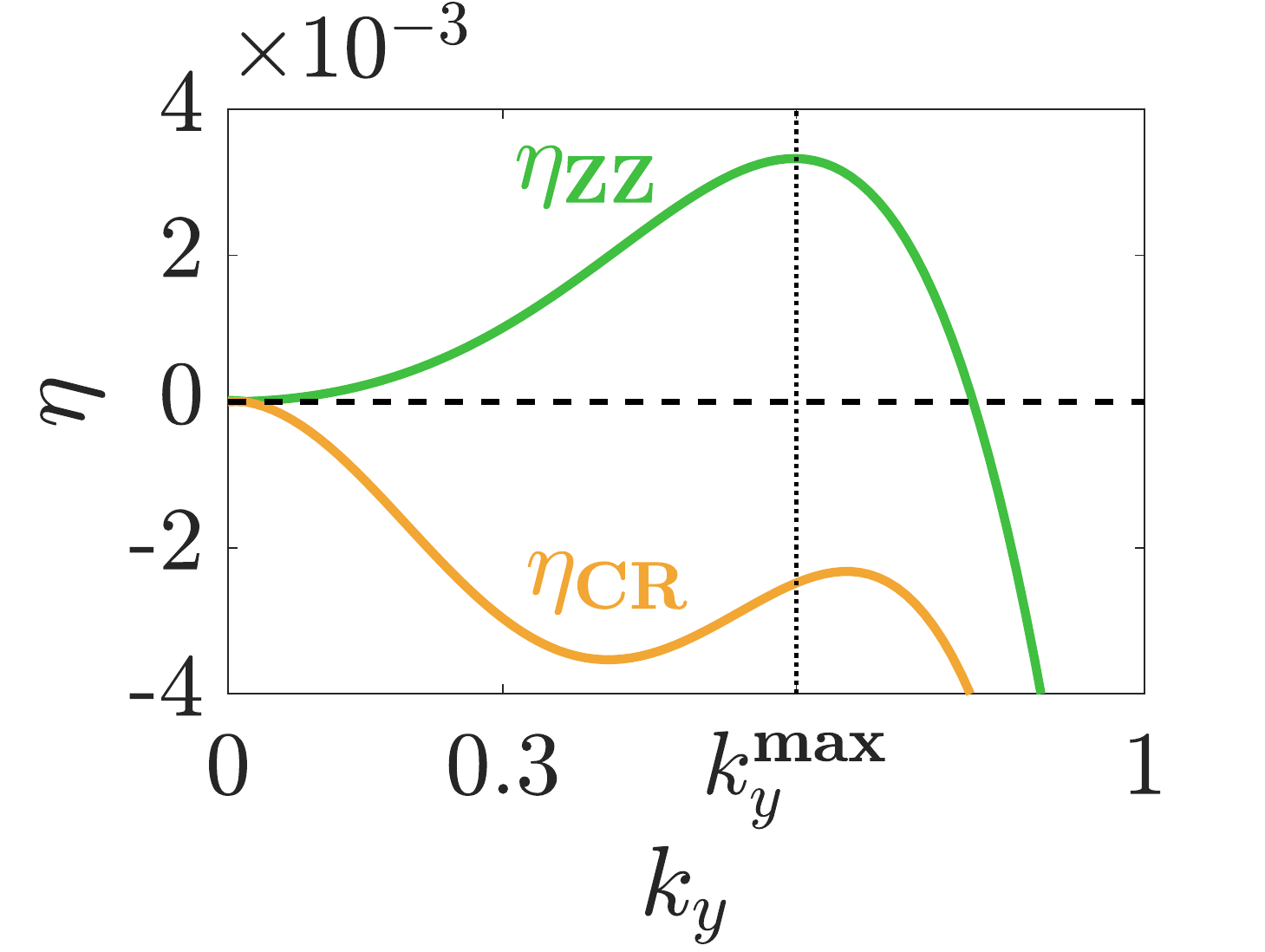}&
			(b)\hspace{3mm}\includegraphics[height=0.18\textheight,valign=t]{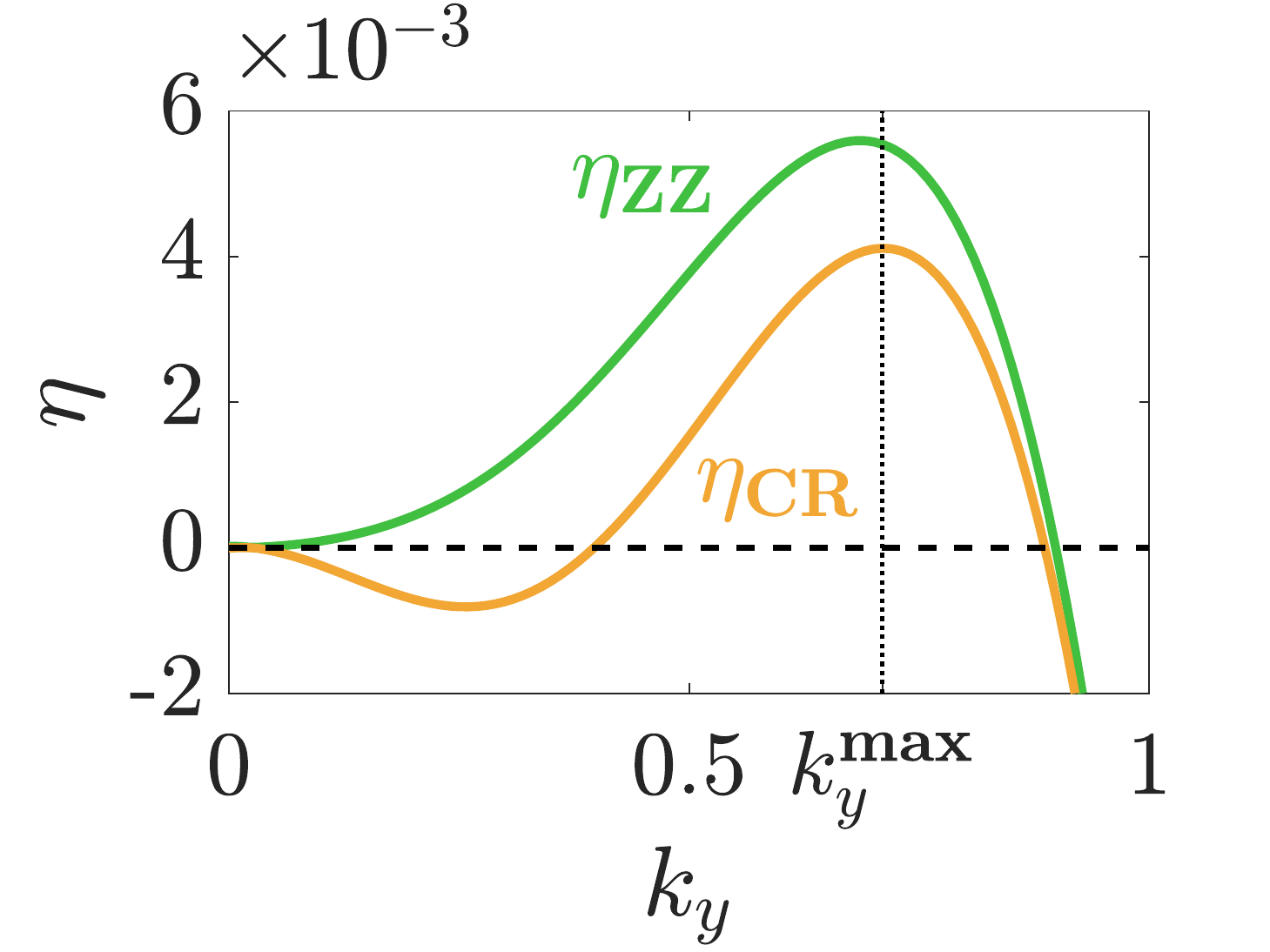}\vspace{5mm}\\
			(c)\includegraphics[height=0.18\textheight,valign=t]{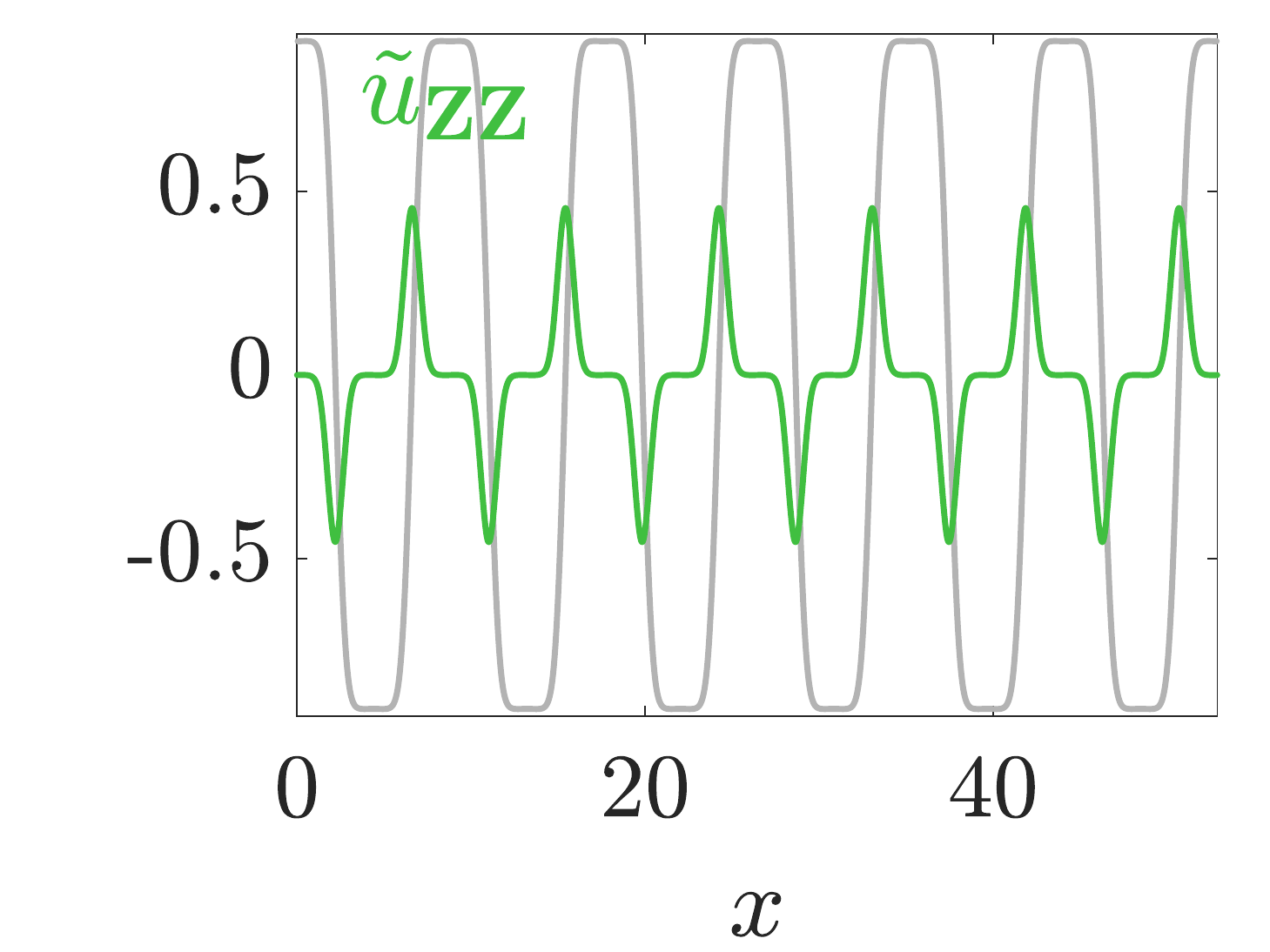}&
			(d)\includegraphics[height=0.18\textheight,valign=t]{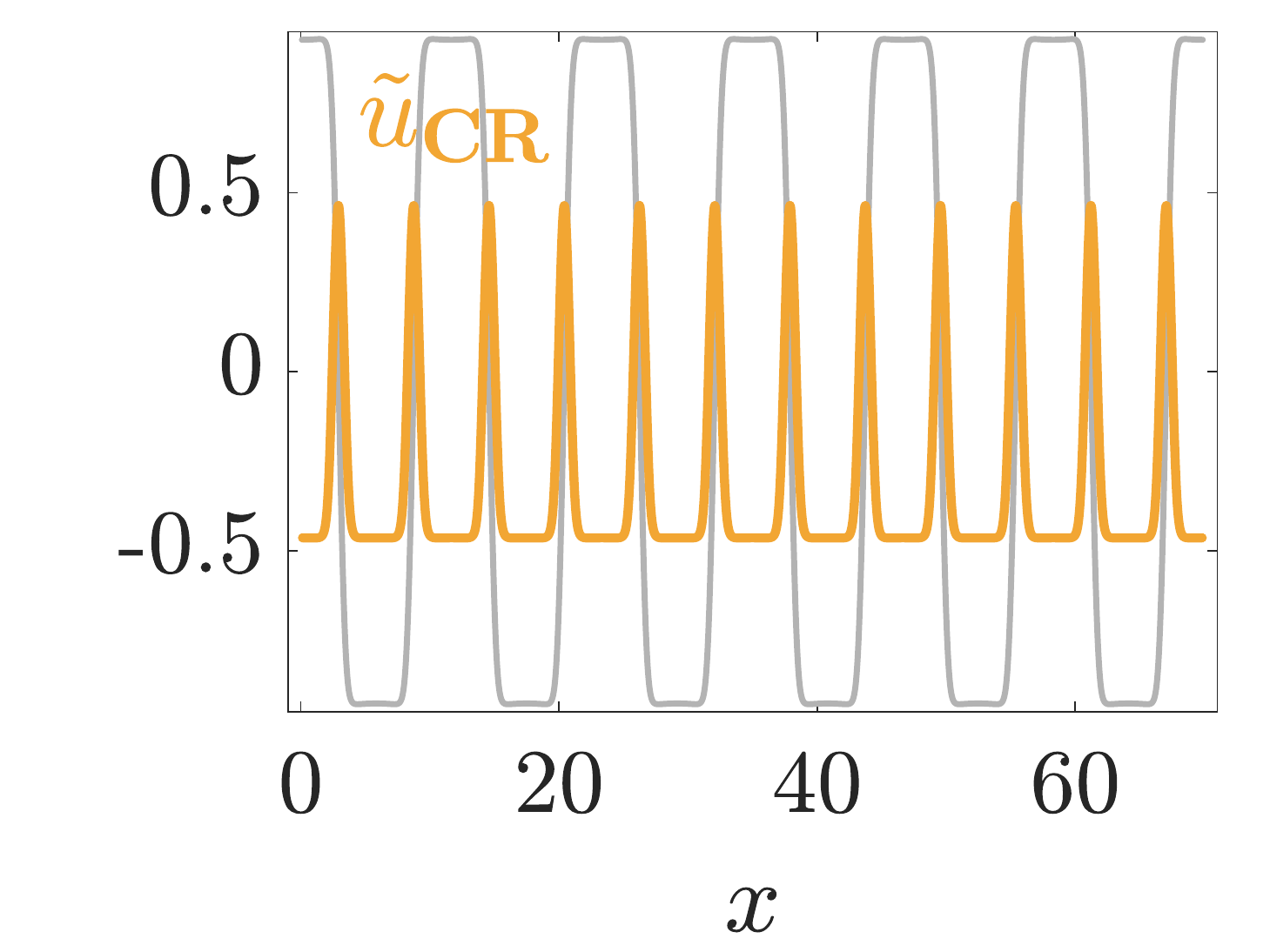}
		\end{tabular}
		\caption{(a) Dispersion relations showing the growth rates of the unstable ZZ ($\eta_\text{ZZ}$, green) and the stable CR ($\eta_\text{CR}$, orange) modes at $\xi=2.65$; at $\ky$, $\eta_\text{ZZ}$ is maximal. (b) Same as (a) but for $\xi=2.8$ where also the CR mode is unstable; here, $\ky$ marks the maximum of $\eta_\text{CR}$. (c) Normalized eigenfunctions of the ZZ instability, $\tilde{u}_\text{ZZ}(x)$ centered around $u=0$ and rescaled to one half the amplitude of the periodic solution, $u_\ell$ (gray line) for $\xi=2.65$. (d) Same as (c) but for the CR eigenfunction $\tilde{u}_\text{CR}(x)$ (orange) at $\xi=2.8$.
			Parameters: $G=8$ and other parameters as in Fig.~\ref{fig:pre.bif}.}\label{fig:pre.evp}
	\end{figure}
	
	In Fig.~\ref{fig:pre.evp}, we show two numerical realizations that produce 
	the instabilities schematically depicted in Fig.~\ref{fig:pre.zzcr}, 
	for different MP well depths $\xi=2.65$ (in (a)) and $\xi=2.8$ (in (b)) while keeping illumination fixed, $G=8$. The dispersion relations ($\eta_\text{ZZ}$ and $\eta_\text{CR}$) indicate that while in both cases the ZZ (odd symmetry) mode is unstable for $G>G_\text{ZZ}$, the CR (even symmetry) mode becomes unstable (with $\eta_\text{CR}(\ky)>0$) only above $G=G_\text{CR}$; in (c,d) we also show the corresponding $\tilde{u}$ component of the eigenfunctions for the ZZ or the CR modes at $\ky$. 
	\begin{figure*}[t!]
		\includegraphics[width=1\textwidth]{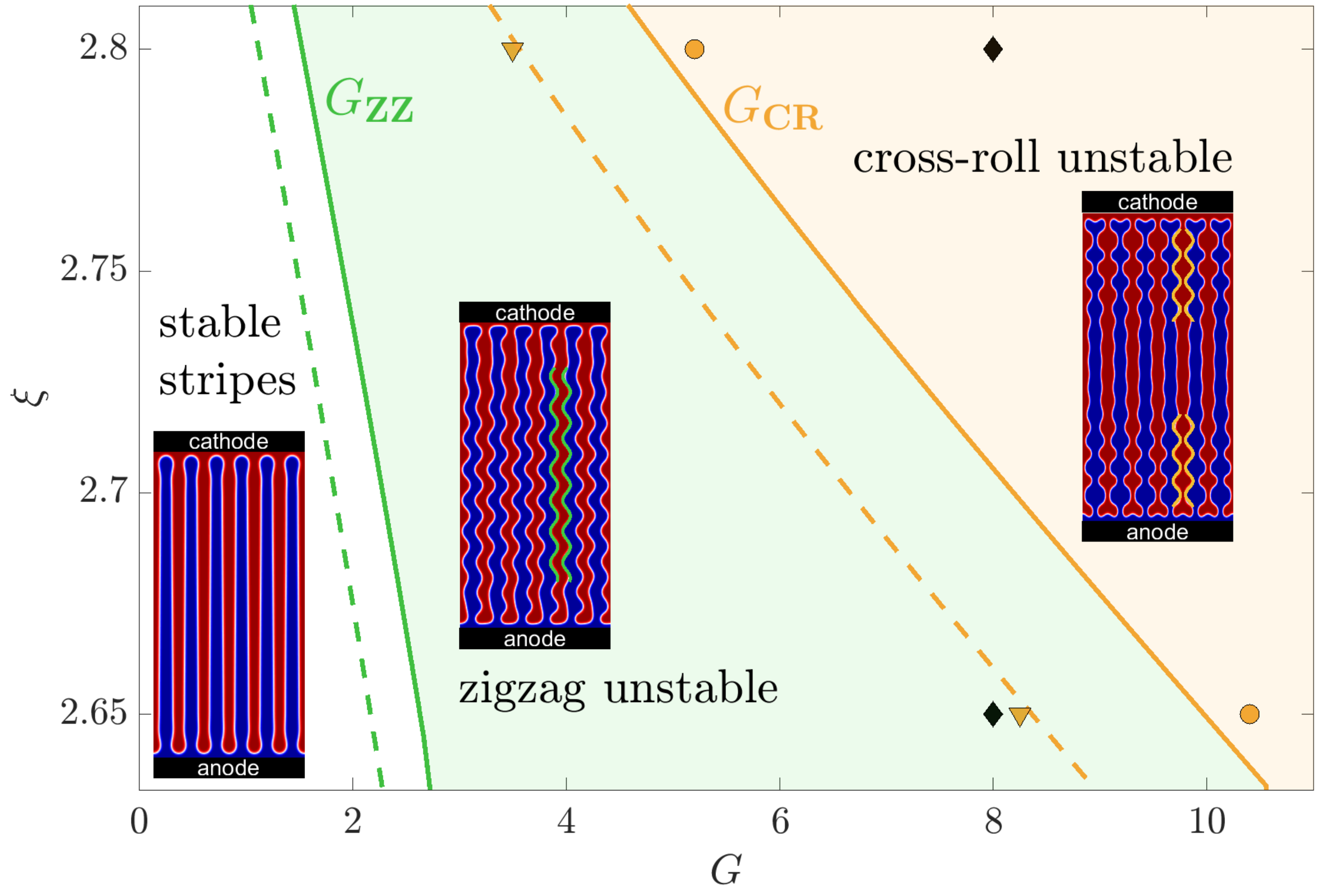}
		\caption[Zigzag and cross-roll unstable regions and numerical simulations]{Parameter plane $(G,\xi)$, corresponding to illumination ($G$) and mixing energy ($\xi$), showing the instability regions of stripes. Solid/dashed lines denote the onsets $G_\text{ZZ}$ and $G_\text{CR}$ for the symmetric/asymmetric D-A ratio ($\langle u \rangle=0$ / $\langle u \rangle=0.1$) as obtained from~\eqref{eq:pre.evp} while `$\bullet$' and `$\blacktriangledown$' correspond to onsets obtained by numerical integration of~\eqref{eq:pre.model} with periodic boundary conditions in the $x$ direction and charge flux in the $y$ direction~\cite{shapira2019pattern} (see text for details). The insets show snapshots of $u$ obtained by numerical integration of~\eqref{eq:pre.model}: left inset is asymptotic solution at $(G,\xi)=(2,2.65)$, middle inset for $(G,\xi)=(8,2.65)$ and $t=2800$ as indicated by the bottom diamond, and right inset for $(G,\xi)=(8,2.8)$ and $t=1200$ as indicated by the top diamond. The green and orange envelope lines in the middle and the right insets represent the ZZ and the CR modes, respectively, as obtained by eigenvalue analysis~\eqref{eq:pre.evp}: the modes are parameterized as $x=(4\mp1/4)\ell_c-\varepsilon\cos(\ky y)$ with $\varepsilon=0.15$ and $\ky=0.62$ as indicated in Figure \ref{fig:pre.evp}(a) and $x=(4\mp1/4)\ell_c\pm\varepsilon\cos(\ky y)$ with $\varepsilon=0.15$ and $\ky=0.71$ as indicated in Figure \ref{fig:pre.evp}(b), respectively. The simulations were performed on a domain $\Omega=[0,6\ell_c]\times[0,100]$: for ZZ with $\ell_c\approx8.81$ in $x$ at $\xi=2.65$ and for CR with $\ell_c\approx11.66$ at $\xi=2.8$. 
			Other parameters are as in Figure \ref{fig:pre.bif}.}\label{fig:pre.onsets}
	\end{figure*}
	
	{We generalize the results in a parameter plane spanned by ($G,\xi$) (see Fig.~\ref{fig:pre.onsets}) and show that a similar trend persists (dashed lines) also for the asymmetric donor-acceptor ratio. The instability \textit{onsets} are defined such that the maximal growth rate becomes positive, i.e. when $\eta(\ky)>0$, at $G_\text{ZZ}$ and $G_\text{CR}$, respectively. This implies degeneracy for $G>G_\text{CR}$, the onset of CR mode, a region in which a competition between bending and pinching of stripes should be expected (even though the ZZ mode has a larger growth rate, $\eta_\text{ZZ}>\eta_\text{CR}$). Next, we show that this degeneracy is destroyed once we allow passage of current through boundaries in the $y$ direction, i.e., physical boundary conditions.} 
	
	\subsection{Realization of instability modes in the presence of charge outflux}
	
	{Although in the above analysis we used non-physical boundary conditions as we did not allow charge flux through the boundaries in $y$ direction, the results provide a good guiding for realistic charge-flux boundary conditions~\cite{Buxton2006}. We validate these results by performing direct numerical simulations using~\eqref{eq:pre.model} with outflux of charges through the $y$ boundaries, assuming that these represent the charge collectors/electrodes~\cite{shapira2019pattern}:
		\begin{eqnarray*}
			\hspace{-5mm}\left(J^u_y,J^\chi_y,J^p_y,J^n_y,\phi \right)\big\vert_{y=0}&=&\left(0,0,-D_p p\frac{\partial \phi}{\partial y},0,\frac{V}{2}\right),\\ \left(J^u_y,J^\chi_y,J^p_y,J^n_y,\phi \right)\big\vert_{y=L_y}&=&\left(0,0,0,-D_n n\frac{\partial \phi}{\partial y},-\frac{V}{2}\right),
		\end{eqnarray*}
		where $V=0$ is a fixed voltage under short circuit conditions, and the fluxes (in their dimensionless forms) are
		\begin{equation*}
			{\mathbf J}^u=D_u(1-u^2)\nabla\frac{\delta\mathcal{E}}{\delta u},~~{\mathbf J}^\chi=\chi\nabla\frac{\delta\mathcal{E}}{\delta\chi},~~{\mathbf J}^p=D_p\,p\nabla\frac{\delta\mathcal{E}}{\delta p},~~{\mathbf J}^n=D_n\,n\nabla\frac{\delta\mathcal{E}}{\delta n}.
		\end{equation*}
		In the $x$-direction we employ periodic BC for all fields.}
	
	At low illumination values, $G<G_\text{ZZ}$, we find that the stripes are stable (left inset in Fig.~\ref{fig:pre.onsets}). In the region $G_\text{ZZ}<G<G_\text{CR}$, the stripes are unstable only to ZZ, which as can be expected develops in the bulk (middle inset in Fig.~\ref{fig:pre.onsets}). The agreement with the linear analysis is excellent and reproduces similar wavenumber $\ky=0.62$, as shown by the green curves in the middle inset. 
	
	In contrast, for $G>G_\text{CR}$ the primary instability now develops near the $y$ boundaries and is of a cross-roll mode type (right inset in Fig.~\ref{fig:pre.onsets}). Consequently, the charge-flux boundary conditions break the degeneracy of the ZZ and the CR modes by enhancing the latter. Nevertheless, the results {are in agreement} 
	with the linear analysis (see orange lines near the boundaries) for both the developed wavenumbers and the onsets (as shown by the dots (symmetric case) and inverted triangles (asymmetric case)). Consequently, these results indicate that decreasing $\xi$ and thus, pronouncing the mixing energy towards a triple well, shifts the instability onsets to higher $G$ values. The latter in turn, suggests that the OPV will become less susceptible to deformation modes that enhance morphological degradation, in particular the dangerous CR instability.
	
	\section{Discussion}
	Following recent highlights of a three-phase (donor/mixed/acceptor) bulk-heterojunction (BHJ) in organic photovoltaics (OPV)~\cite{dkhil2017toward,ma2014quantification,zhou2019hierarchical,wang2017ternary}, we used a mean-field approach~\cite{shapira2019pattern} to identify the role of the intermediate mixed-phase on morphological changes. Under illumination the model is driven out of equilibrium so that stripe morphology may arise (Fig.~\ref{fig:pre.bif}).  In contrast, under dark conditions the system evolves solely by coarsening~\cite{Buxton2006,dkhil2017toward}. From a mathematical point of view, stripe morphology arises due to a finite wavenumber instability~\cite{shapira2019pattern} that is possible only under illumination and whose nature is effected by the order parameter and the exciton/electron/hole fields (see system~\eqref{eq:pre.model}). We focus on and distinguish between two generic transverse instabilities of donor-acceptor stripes in 2D (distinctly from the formation of stripes by phase separation) with symmetric and asymmetric compositions (as summarized in Fig.~\ref{fig:pre.onsets}): the bending (zigzag mode) and the pinching (cross-roll mode). {The pinching mode is characterized by high mixing energy whereas at low mixing energies bending of the donor/acceptor domains is favored. We emphasize that the time scale separation between the morphological (material) changes and charge dynamics is of several orders of magnitude so that our results indicate only the initial trend and not necessarily convergence to a final state, but the further evolution, in reality, is extremely slow. Furthermore, the slow time evolution of the material lowers the sensitivity of the OPV to finite-amplitude perturbations, thus, the effect for example of sudden changes in illumination is negligible.}
	
	{Although we limited our analysis to 2D, standard theory shows that the pinching mode may also lead to discontinuous and isolated domains in 3D~\cite{yu1994mechanism,kolmychkov2005identification,fedoseev20103d,uecker2020snaking} and thus, in OPV loss of current to the electrodes that cause operation failure. This phenomenon resembles the so-called pearling of cylindrical threads~\cite{tsafrir2001pearling,nelson1995dynamical,sinha2013electric,chaieb1998spontaneous,nguyen2005surface}. Moreover, according to numerical simulations, relatively large D/A volumes of BHJ, are more susceptible to transverse instabilities since the intermediate phase does not suppress transverse front instabilities that arise due to curvature effects as in bistable systems~\cite{goldstein1996interface,Yochelis2004,Hagberg2006,kolokolnikov2007spot}, i.e., in the direction that is parallel to the electrodes. This is consistent with the diffusion length of about tens of nanometer size of the BHJ~\cite{jorgensen2012stability,ma2014quantification}.} 
	
	Consequently, we showed that the qualitative significance of three-phase BHJ goes beyond inhibition of phase separation~\cite{dkhil2017toward}, as it may have \textit{tailoring by demand} properties that can be controlled by the composition of the mixed-phase via donor-acceptor choices: {by decreasing the mixing energy parameter~$\xi$, 
		the instability onsets are shifted to higher illumination values~$G$.} This degree of control is absent or less sequential in two-phase OPV. We believe that our results may assist in the future design of long-lasting OPV, consisting of three-phase BHJ. In a broader context, our results should apply to other systems in physicochemical systems that exhibit phase separation~\cite{emmerich2008advances,dewitt2018phase} and can be driven out of equilibrium, in particular in ion-intercalated renewable batteries that depend on reversible phase exchanges in charge/discharge cycling~\cite{kaufman2019understanding,balakrishna2019phase,van2020rechargeable}, such as Li~\cite{tang2010electrochemically,grazioli2016computational,zhao2019review} and Ni~\cite{briggs1971oxidation,barnard1980studies,huggins1994proton} based electrodes.
	
	\begin{acknowledgments}
		The research was
		done in the framework of the Grand Technion Energy Program (GTEP) and of the BGU Energy Initiative Program, and supported by the Adelis Foundation for renewable energy research.
	\end{acknowledgments}

\end{document}